\documentstyle[11pt]{article}
\def\NPB#1#2#3{Nucl. Phys. B{#1} (19#2) #3}
\def\PLB#1#2#3{Phys. Lett. B{#1} (#2) #3}

\def\PRD#1#2#3{Phys. Rev. D{#1} (19#2) #3}

\def\MODA#1#2#3{Mod. Phys. Lett.  {#1} (19#2) #3}

\def\nn{\nonumber \\ &&}

\def\no{\nonumber \\ }
\def\be{\begin{eqnarray}}
\def\ee{\end{eqnarray}}
\def\ben{\begin{equation}}
\def\een{\end{equation}}

\begin{document}

\makeatletter \@addtoreset{equation}{section}
\makeatother
\renewcommand{\theequation}{\thesection.\arabic{equation}}
\begin{flushright}
HUE-02/1 \\
\end{flushright}
\bigskip
\bigskip
\begin{center}
{\LARGE{Crosscap states and Boundary states in $D=4,N=1$,type-IIB 
Orientifold Theories}  \\}
\bigskip
H. Kataoka 
\footnote{kataoka@yukawa.kyoto-u.ac.jp} 
and Hikaru Sato
\footnote{hikaru@sci.hyogo-u.ac.jp} 
\\
\bigskip
{\small \it
Department of Physics,
Hyogo University of Education,\\
Yashiro-cho, Hyogo 673-1494, JAPAN}
\end{center}
\begin{abstract}
We construct boundary state and crosscap state in $D=4,N=1$ 
type-IIB $Z_N$ orientifold 
and investigate properties of amplitude.
We find that the boundary state of a cylinder is different from the 
boundary state of a M\"{o}bius strip.
Using these states,
we find that
amplitudes 
do not factorize
in $Z_N(N=$even) orientifold.
Tadpole divergence remain 
in $Z_4, Z_8, Z'_8$ and $Z'_{12}$
 model due to volume dependence of
boundary and crosscap state.
On the other hand
the amplitude of $Z_3$ and $Z_7$ orientifolds factorize 
so that we obtain the gauge groups of the model 
by employing the massless tadpole cancellation condition.
\end{abstract}
\section{Introduction}
The boundary state formalism was introduced to interprete open string
 amplitudes in terms of closed string states.
This could be valuable in understanding the relationship between 
closed and open strings which is one of the central problems in 
uncovering the underlying symmetry of string theory.
It serves very useful in analyzing the spectrum of models that do 
not have an obvious geometrical interpretation such as orbifold 
with discrete torsion \cite{vaf}.
Recently, boundary state formalism in conformal field theory is 
explored \cite{dia}\cite{mer}\cite{bil}.
The boundary state formalism is a powerful framework for studying 
D-branes \cite{car}\cite{cp},
and useful for
computing D-brane tensions and cylinder amplitudes as well as in 
looking for the gravity counterparts of D-branes.
The structure of $ D=4,N=1,$ type-IIB orientifold is explored, and 
conditions for tadpole cancellation in type-IIB $Z_N$ orientifold 
have been presented \cite{ber}.
However, in spite of importance to understand the relationship 
between closed and open string, 
the relation is not investigated in $ D=4,N=1,$ type-IIB orientifold.

In this paper, we derive crosscap states 
 and boundary states for 
D-branes at a fixed point in $ D=4,N=1,$ type-IIB $Z_N$ orientifold 
theory.
We find  that a boundary state in M\"{o}bius strip is different from 
that in cylinder in general.
And we investigate the structure of factorization in $Z_N$ orbifolds.
Even though amplitude factorizes in $D=10$ models in 
general, amplitude does not factorize in general
in $D=4, Z_N(N=$even) models.
In section 3 we consider momentum and winding modes of boundary states and 
crosscap states.
The tadpole divergence remains 
in $Z_4,
 Z_8 ,Z'_8$ and $Z'_{12}$ model.
Using tadpole cancellation condition, we derive gauge groups 
in $Z_3,Z_7$ models.
\section{Construction of crosscap and boundary states}
At first, we construct a crosscap state $|C\rangle$ in
$D=4,N=1,$ type-IIB $Z_N$ orientifold  
model.
We summarize in Table 1 the $Z_N$ action that 
leads to $N=1$ supersymmetry .

The mode expansion of a closed string state reads
\ben
X^\mu(\sigma_1, \sigma_2)=x^{\mu}+l^2p^\mu\sigma_1+\frac{i}{2}l
\sum_{n\ne0}(\frac{1}{n}\alpha^\mu_ne^{-2i\pi n(\sigma_2-\sigma_1)}
+\frac{1}{n}\tilde{\alpha}^\mu_ne^{-2i\pi n(\sigma_2+\sigma_1)}).
\een
We introduce the complex coordinate,
\ben
Y^i=X^{2i+2}+iX^{2i+3},\quad\bar{Y}^i=X^{2i+2}-iX^{2i+3}, \quad 
(i=1,2,3).
\een
The $Z_N$ orientifold model has $k$-twisted sectors of the closed 
string.
The boundary condition for the Klein bottle is
\ben
Y^i(\sigma_1+1, \sigma_2)=e^{2i\pi kv_i}Y^i (\sigma_1,\sigma_2)
\hspace{2cm}(i=1,2,3),
\een
with a fundamental domain $0\le\sigma_1<1$ and $0\le\sigma_2<1$ in
 Figure 1.
By using another fundamental domain $0\le\sigma_1<\frac{1}{2}$ 
and $0\le\sigma_2<2$, 
the boundary condition for the Klein bottle is
\ben
Y^i(0, \sigma_2+2)=e^{4i\pi kv_i}Y^i(0, \sigma_2), \hspace{1cm}
Y^i(\frac{1}{2}, \sigma_2+2)=Y^i(\frac{1}{2}, \sigma_2).
\een
It means $Y^i$ at $\sigma_1=0$ has $2k$-twisted 
sectors,
while $Y^i$ at $\sigma_1=\frac{1}{2}$ is not twisted.  

The state produced from the vacuum by crosscap is determined 
up to normalization 
by crosscap conditions on the fields.
Here we define $\tau=2\sigma_1, \sigma=\frac{1}{2}\sigma_2$.
And we denote $Y^i$ at $\sigma_1=0$, $Y^i_0(\sigma)$,
and at $\sigma_1=\frac{1}{2}$, $Y^i_{\frac{1}{2}}(\sigma)$.
$Y^i_{0,\frac{1}{2}}(\sigma)$ has $2m$-twisted sectors.
Then 
\ben
Y^i_0(\sigma+1)=e^{4i\pi (kv_i+mv_i)}Y^i_0(\sigma),
\hspace{.7cm}
Y^i_{\frac{1}{2}} (\sigma+1)=e^{4i\pi mv_i} Y^i_{\frac{1}{2}}
 (\sigma).
\een
We denote $2k$-twisted, $2m$-twisted crosscap state by 
$|c,2kv_i,2mv_i\rangle^i$,
where $v_i$ 
is a component of the twist vector $(v_1,v_2,v_3)$ in Table 1.
The crosscap state conditions read
\be
\Big(Y^i_0(\sigma+\frac{1}{2})-e^{2i\pi (kv_i+mv_i)}Y^i_0
(\sigma)\Big) 
|c,2kv_i,2mv_i\rangle^i &=&0,  \no
\Big(\partial_{\tau}Y^i_0(\sigma+\frac{1}{2})+e^{2i\pi (kv_i+mv_i)}\partial_{\tau}Y^i_0(\sigma)\Big) |c,2kv_i,2mv_i
\rangle^i &=&0,  \no
\Big(Y^i_{\frac{1}{2}}(\sigma+\frac{1}{2})- e^{2i\pi mv_i} 
Y^i_{\frac{1}{2}} (\sigma)\Big)|c,0,2mv_i\rangle^i &=&0 , \no
\Big(\partial_{\tau} Y^i_{\frac{1}{2}} (\sigma+\frac{1}{2})+ 
e^{2i\pi mv_i}\partial_{\tau} Y^i_{\frac{1}{2}} (\sigma)\Big)
|c,0,2mv_i\rangle^i &=&0 
\label{beta}. 
\ee
By solving the conditions(\ref{beta}), the crosscap states are  
\be
|c, 2kv_i ,2mv_i\rangle^i
&=&\prod_{n=1}  
\exp[\frac{- e^{-i\pi (n-1)}\bar{\beta}^i_{-n+1-2(k+m)v_i}
\tilde{\beta}^i_{-n+1-2(k+m)v_i}}
{n-1+2(k+m)v_i}]\nn
\exp[\frac{- e^{i\pi n}\beta^i
_{-n+2(k+m)v_i}\tilde{\bar{\beta}}^i_{-n+2(k+m)v_i}}
{n-2(k+m)v_i}] |0\rangle, 
\ee
where $|0\rangle $ is the Fock vacuum and
$\beta_n^i$ are oscillator modes of $Y^i$ defined by
\be
\beta_n^i=&\alpha_n^{2i+2}+i\alpha_n^{2i+3},\quad
\bar{\beta}_n^i&=\alpha_n^{2i+2}-i\alpha_n^{2i+3}, \quad\no
\tilde{\beta}_n^i=&\tilde{\alpha}_n^{2i+2}+i\tilde{\alpha}_n^{2i+3},
 \quad
\tilde{\bar{\beta}}_n^i&=\tilde{\alpha}_n^{2i+2}-
i\tilde{\alpha}_n^{2i+3} 
\quad(i=1,2,3).
\ee

Next we consider fermionic parts.
Fermionic parts of the closed string are
\ben
\psi^\mu(\sigma_1, \sigma_2)=\sqrt{2\pi}\sum_r\psi^\mu_re^{-2i\pi r
(\sigma_2-\sigma_1)},\quad
\tilde{\psi}^\mu(\sigma_1, \sigma_2)=\sqrt{2\pi}\sum_r
\tilde{\psi}^\mu_r
e^{-2i\pi r(\sigma_2-\sigma_1)}.
\een
Similar conditions as the bosonic case lead to the crosscap states,
\be
|c,2kv_i,2mv_i\rangle^i 
 &=&
\prod_{n=1}\exp[i\eta e^{-i\pi n}\bar{\lambda}^i
_{-n+1-2(k+m)v_i}\tilde{\lambda}^i_{-n+1-2(k+m)v_i} ]\nn
\exp[-i\eta e^{-i\pi n}\lambda^i_
{-n+2(k+m)v_i}\tilde{\bar{\lambda}}^i
_{-n+2(k+m)v_i} ] |0,\eta \rangle.
\ee
where $|0,\eta \rangle  $ is a usual Ramond-Ramond vacuum 
$\eta=\pm 1$
and
\be
\lambda^i_r=&\psi^{2i+2}_r
+i\psi^{2i+3}_r,\hspace{1cm}
\bar{\lambda}^i_r=&\psi^{2i+2}_r
-i\psi^{2i+3}_r,\no
\tilde{\lambda}^i_r=&\tilde{\psi}^{2i+2}_r
+i\tilde{\psi}^{2i+3}_r, \hspace{1cm}
\tilde{\bar{\lambda}}^i
=&\tilde{\psi}^{2i+2}_r-i\tilde{\psi}^{2i+3}_r.
\ee
By combining the bosonic and fermionic contributions,
the crosscap state are defined by 
\be
|C,2k\rangle&=&\sum^{N-1}_{m=0} N_{c,2k,2m}|c,2k,2m\rangle,\no
|c,2k,2m\rangle
&=& |c,2kv_3,2mv_3\rangle^3|c,2kv_2,2mv_2\rangle^2
|c,2kv_1,2mv_1\rangle^1
|c,0,0\rangle^0,
\label{c}
\ee
where $|c,0,0\rangle^0$ denotes the crosscap state for the 
uncompactified $(X^2,X^3)$ plane.
The coefficients $N_{c,2k,2m}$ will be determined by 
tadpole cancellation conditions.
We denote especially $k=0$ crosscap state by $|C\rangle$.
 
Next we consider a boundary state $|B\rangle$.
Boundary state of cylinder has already constructed in 
ref.\cite{dia}.
The results are
\be
|b,0,mv_i\rangle^i_{DD}
&=&\prod_{n=1}
\exp[\frac{\bar{\beta}^i_{-(n-1)-mv_i}\tilde{\beta}^i
_{-(n-1)-mv_i}}
{n-1+mv_i}]
\exp[\frac{\beta^i_{-n+mv_i}\tilde{\bar{\beta}}^i_
{-n+mv_i}}{n-mv_i}
]\no
&&
\prod_{n=1}\exp[i\eta\bar{\lambda}^i_{-n+1-mv_i}\tilde
{\lambda}^i
_{-n+1-mv_i} ] 
\exp[i\eta\lambda^i_{-n+mv_i}\tilde{\bar{\lambda}}^i_{-
n+mv_i} ]|0,\eta 
\rangle,\no
|b,0,mv_i\rangle^i_{NN}
&=&\prod_{n=1} \exp[
\frac{-\bar{\beta}^i_{-(n-1)-mv_i}\tilde{\beta}^i
_{-(n-1)-mv_i}}
{n-1+mv_i}]
\exp[
\frac{-\beta^i_{-n+mv_i}\tilde{\bar{\beta}}^i_
{-n+mv_i}}{n-mv_i}] \no
&&
\prod_{n=1}\exp[-i\eta\bar{\lambda}^i_{-n+1-mv_i}\tilde
{\lambda}^i
_{-n+1-mv_i} ]
\exp[-i\eta\lambda^i_{-n+mv_i}\tilde{\bar{\lambda}}^i
_{-n+mv_i} ] 
|0,\eta \rangle. \no 
\ee
Here subscripts DD and NN mean Dirichlet 
and Neumann boundary conditions
in $X^{2i+2},X^{2i+3}$ $(i=1,2,3)$
directions, respectively.
D9-branes has Neumann boundary conditions in $X^{\mu}(\mu=1,
\dots 9)$ directions.
D5-branes has Neumann boundary conditions in 
$X^{\mu}(\mu=1,2,3,8,9)$ directions 
and Dirichlet boundary conditions in $X^{\mu}(\mu=4,5,6,7)$ 
directions.
Hence the boundary states $|B^c\rangle$ for cylinder on D5/D9-branes 
that include the Chan-Paton factors 
are defined as
\be
|B^c\rangle_{p}&=&\sum^{N-1}_{m=0} N^c_{b,0,m,p}|b,0,m\rangle_{p}
(Tr \gamma_{m,p}) \hspace{1cm}
(p=5,9),\no
|b,0,m\rangle_{5}
&=&|b,0,mv_3\rangle^3_{NN}|b,0,mv_2\rangle^2_{DD}
|b,0,mv_1\rangle^1_{DD}|b,0,0\rangle^0_
{NN},\no
|b,0,m\rangle_{9}
&=&|b,0,mv_3\rangle^3_{NN}|b,0,mv_2\rangle^2_{NN}
|b,0,mv_1\rangle^1_{NN}
|b,0,0\rangle^0_{NN}.
\label{bc}
\ee
Coefficients $N^c_{b,0,m,p}$ will be determined by 
tadpole cancellation condition.

Next we consider M\"{o}bius strip.
We have M\"{o}bius strip boundary conditions, 
\be
Y^i(1, \sigma_2)=Y^i(0, \sigma_2+1),&\hspace{1cm}&
\partial_{1}Y^i(0, \sigma_2)= 0\no
Y^i(\frac{1}{2}, \sigma_2)= 
Y^i(\frac{1}{2}, \sigma_2+1),&\hspace{1cm}&
\partial_{1}Y^i(\frac{1}{2}, 
\sigma_2)=-\partial_{1}Y^i(\frac{1}{2},
 \sigma_2+1)
\ee
Similar consideration like Klein bottle leads
to the conclusion that 
the boundary $Y^i_0(\sigma)$ and the crosscap
$Y^i_{\frac{1}{2}}(\sigma)$ 
do not possess $k$-twisted sectors but have $2m$-twisted sectors:
\ben
Y^i_0(\sigma+1)=e^{4i\pi mv_i}Y^i_0(\sigma),\quad
Y^i_{\frac{1}{2}} (\sigma+1)=e^{4i\pi mv_i} Y^i_
{\frac{1}{2}} (\sigma).
\een
We denote $2m$-twisted crosscap state by $|c,0,2mv_i \rangle^i$,
 and 
$2m$-twisted boundary state by $|b,0,2mv_i\rangle^i$.
The crosscap states $|c,0,2mv_i\rangle^i$ are same as those 
for the Klein bottle.
But boundary state $|b,0,2mv_i\rangle^i$ which is 
$2m$-twist is  different 
from boundary state in cylinder $|b,0,mv_i\rangle^i$ which is 
$m$-twist.
We determine 
boundary states $|B^M\rangle$ for M\"{o}bius strip that include 
the Chan-Paton factors.
\be
|B^M\rangle_{p}&=&\sum^{N-1}_{m=0} N^M_{b,0,2m,p}|b,0,2m\rangle_{p}
(Tr \gamma_{2m,p}),\hspace{1cm}
(p=5,9),\no
|b,0,2m\rangle_{5}
&=&|b,0,2mv_3\rangle^3_{NN}|b,0,2mv_2\rangle^2_{DD}
|b,0,2mv_1\rangle^1_{DD}|b,0,0\rangle^0_
{NN},\no
|b,0,2m\rangle_{9}
&=&|b,0,2mv_3\rangle^3_{NN}|b,0,2mv_2\rangle^2_{NN}
|b,0,2mv_1\rangle^1_{NN}
|b,0,0\rangle^0_{NN}.
\label{bm}
\ee

By using crosscap state (\ref{c}), 
boundary states (\ref{bc}) and (\ref{bm})
and Cardy's condition \cite{bil}\cite{car},
 the amplitudes for Klein bottle, 
cylinder and M\"{o}bius strip are summarized as
\be
&&{\cal K}=\sum_I \int dl 
(\langle C_I|e^{-lH}|C_I\rangle+
\langle C_I,N|e^{-lH}|C_I\rangle)
=\frac{V_4}{N}\sum^{N-1}_{m=0}
\int^{\infty}_0\frac{dt}{16t^3}\frac{\tilde{\vartheta}
[^{0}_{\frac{1}{2}}]
_{(2t)}}{\tilde{\eta}_{(2t)}^3}(Z_K+ Z_{KT}),\no
&&{\cal C}_{pq}=\sum_I\int dl _p\langle B^c_I|e^{-lH}|B^c_I\rangle_q
=\frac{V_4}{N}\sum^{N-1}_{m=0}
\int^{\infty}_0\frac{dt}{4t^3}\frac{\vartheta[^0_{\frac{1}
{2}}]_{(t)}}
{\eta^3_{(t)}}Z_{pq},\no
&&{\cal M}_p=\sum_I \int dl _p\langle B^M_I|e^{-lH}|C_I\rangle
=\frac{V_4}{N}\sum^{N-1}_{m=0}
\int^{\infty}_0\frac{dt}{32t^3}\frac{\tilde{\vartheta}
[^0_{\frac{1}{2}}]
_{(2t)}\tilde{\vartheta}[^{\frac{1}{2}}_{0}]_{(2t)}}
{\tilde{\eta}^3_{(2t)}
\tilde{\vartheta}[^0_{0}]_{(2t)}}Z_{p}.
\ee
An index $I$ stands for the fixed point.
Explicit form of $Z_K, Z_{KT} Z_{pq}$ and $Z_p$ are
\be
Z_K& =&
\prod_{i=1}^3
\frac{-2\sin2\pi mv_i 
\tilde{ \vartheta}[^{0}_{2mv_i+\frac{1}{2}}]_{(2t)}}
{\tilde{\vartheta}[^{\frac{1}{2}}_{2mv_i+\frac{1}{2}}]
_{(2t)}} 
N_{c,0,2m}^2,\no
Z_{KT}&=&
\prod_{i=1}^2
\frac{\tilde{\vartheta}[^{\frac{1}{2}}_{2mv_i+\frac{1}{2}}]
_{(2t)}}
{\tilde{\vartheta}[^{0}_{2mv_i+\frac{1}{2}}]_{(2t)}}
\frac{-2\sin2\pi mv_3 \tilde{\vartheta}
[^{0}_{2mv_3+\frac{1}{2}}]_{(2t)}}
{\tilde{\vartheta}[^{\frac{1}{2}}_{2mv_3+\frac{1}{2}}]
_{(2t)}} 
N_{c,N,2m} N_{c,0,2m},\no
Z_{55}&=&
\prod_{i=1}^3\frac{-2\sin\pi mv_i 
\vartheta[^0_{mv_i+\frac{1}{2}}]_{(t)}}{\vartheta
[^{\frac{1}{2}}_{mv_i
+\frac{1}{2}}]_{(t)}}
\sum_I(Tr \gamma_{m,5,I})^2 (N^c_{b,0,m,5})^2,\no
\label{eq:c55}
Z_{99}&=&
\prod_{i=1}^3
\frac{-2\sin\pi mv_i 
\vartheta[^0_{mv_i+\frac{1}{2}}]_{(t)}}{\vartheta
[^{\frac{1}{2}}_{mv_i
+\frac{1}{2}}]_{(t)}}
(Tr \gamma_{m,9})^2 (N^c_{b,0,m,9})^2,\no
Z_{59}&=&
\frac{-2\sin\pi mv_3 
\vartheta[^0_{mv_3+\frac{1}{2}}]_{(t)}}{\vartheta
[^{\frac{1}{2}}_{mv_3
+\frac{1}{2}}]_{(t)}}
\prod_{i=1,2}
\frac{\vartheta[^{\frac{1}{2}}_{mv_i+\frac{1}{2}}]_{(t)}}
{\vartheta[^{0}
_{mv_i+\frac{1}{2}}]_{(t)}}
 (Tr \gamma_{m,9})\sum_I(Tr \gamma
_{m,5,I}) N^c_{b,0,m,9}N^c_{b,0,m,5},\no
Z_9&=&
\prod_{i=1}^3
\frac{-2\sin\pi mv_i 
\tilde{\vartheta}[^{0}_{mv_i+\frac{1}{2}}]_{(2t)}\tilde
{\vartheta}
[^{\frac{1}{2}}_{mv_i}]_{(2t)}}
{\tilde{\vartheta}[^{0}_{mv_i}]_{(2t)}\tilde{\vartheta}
[^{\frac{1}{2}}
_{mv_i+\frac{1}{2}}]_{(2t)}}
(Tr \gamma_{2m,9}) N^M_{b,0,2m,9}N_{c,0,2m},\no
Z_5&=&
\frac{-2\sin\pi mv_3
 \tilde{\vartheta}[^{0}_{mv_3+\frac{1}{2}}]_{(2t)}\tilde
{\vartheta}
[^{\frac{1}{2}}_{mv_3}]_{(2t)}}
{\tilde{\vartheta}[^{0}_{mv_3}]_{(2t)}\tilde{\vartheta}
[^{\frac{1}{2}}
_{mv_3+\frac{1}{2}}]_{(2t)}}\no
&&\prod_{i=1,2}
\frac{2\cos\pi mv_i
 \tilde{\vartheta}[^{\frac{1}{2}}_{mv_i+\frac{1}{2}}]
_{(2t)}\tilde
{\vartheta}[^0_{mv_i}]_{(2t)}}
{\tilde{\vartheta}[^{\frac{1}{2}}_{mv_i}]_{(2t)}\tilde
{\vartheta}
[^0_{mv_i+\frac{1}{2}}]_{(2t)}}
\sum_I(Tr \gamma_{2m,5,I}) N^M_{b,0,2m,5}N_{c,0,2m}.
\label{eq:m5}
\ee
Note that $Z_{KT}$ does vanish for $Z_N($odd $N$).
Comparing (\ref{eq:m5}) with 
the amplitudes of the one-loop vacuum diagram 
\cite{ald}, 
we obtain the 
following relations,:
\be
(N^c_{b,0,m,5})^2= &(N^c_{b,0,m,9})^2&= 
N^c_{b,0,m,5}N^c_{b,0,m,9}=\frac{1}{32\pi^4},\no
N_{c,0,2m}^2=& N_{c,0,2m}N_{c,N,2m}&=\frac{1}{2\pi^4},\no
N_{c,0,2m}N^M_{b,0,2m,9}=& N_{c,0,2m}N^M_{b,0,2m,5}
&=\frac{-1}{4\pi^4}.
\label{N}
\ee

As we have discussed, 
boundary states in 
M\"obius strip and cylinder 
are different;
$|B^c\rangle_{p}$ is the sum of $m$-twisted states 
but $|B^M\rangle_{p}$ is the sum of
$2m$-twisted states.
The amplitude for the M\"{o}bius strip is
\ben
\langle B^M| e^{-lH}|C\rangle
=\sum_{m=0}^{N-1}N_{b,0,2m}^M N_{c,0,2m}\langle b,0,2m| 
e^{-lH}|c,0,2m\rangle(Tr\gamma_{2m}).
\label{Mm}
\een
When we require a factorization of 
the total amplitude
$(\langle B^c|+\langle C|)e^{-lH}(|B^c\rangle+|C\rangle)$, 
the M\"{o}bius strip amplitude is expressed using $|B^c\rangle$ as 
\ben
\langle B^c| e^{-lH}|C\rangle+\langle C| e^{-lH}
|B^c\rangle\no
=2\sum_{m=0}^{N-1}N_{b,0,2m}^c N_{c,0,2m}\langle b,0,2m| 
e^{-lH}|c,0,2m\rangle(Tr\gamma_{2m}).\no
\label{Mc2}
\een
Therefore equality of (\ref{Mm}) and (\ref{Mc2}) 
requires
$N_{b,0,2m}^M=2N_{b,0,2m}^c
\label{Mc}.$
It leads
\be
&&N_{c,0,m}=N_{c,N,m}
=-2N^M_{b,0,m,9}=-2N^M_{b,0,m,5}\no
&&= -4N^c_{b,0,m,5}=-4N^c_{b,0,m,9}
=\pm\frac{1}{\sqrt{2}\pi^2}
\hspace{1cm}(m=\mbox{even}),
\label{Mc4}
\ee
\ben
 N^c_{b,0,m,5}=N^c_{b,0,m,9}
=\pm\frac{1}{4\sqrt{2}\pi^2}
\hspace{1cm}(m=\mbox{odd}).
\label{Mc3}
\een
In the following, we use notation $|B\rangle$ for 
boundary states in M\"obius strip and cylinder
taking (\ref{Mm})(\ref{Mc2})(\ref{Mc}) into consideration.  

Next let us examine the 
factorization of
amplitude.
It is generally believed that amplitude can be factorized. 
However it is not obvious in $N=1,D=4$ type-IIB $Z_N$ orientifold.
In the case of $Z_3,Z_7$ model
\[(_9\langle B|+\langle C|) e^{-lH} (|B\rangle_9 +
|C\rangle )
=_9\langle B| e^{-lH}|B\rangle_9+\langle C| 
e^{-lH}|C\rangle+2_9\langle B| 
e^{-lH}|C\rangle\]
\ben
{\cal K}\sim\langle C| e^{-lH}|C\rangle,\quad
{\cal M}_9\sim2_9\langle B|
 e^{-lH}|C\rangle,\quad
{\cal C}_{99}\sim _9\langle B| e^{-lH}|B\rangle_9
\een
Hence amplitude does factorize.
However in case of $Z_6$,
the would-be factorized amplitude
\ben
(_5\langle B|+_9\langle B|+\langle C|) 
e^{-lH} (|B\rangle_5+|B\rangle_9+|C\rangle )
\een
contains the amplitudes ${\cal C}_{pp}$ and ${\cal M}_p$ 
but not all of the Klein bottle amplitude ${\cal K}$.
Instead, if we include $\langle C,N|$ into the would-be 
factorized amplitude,
\ben
(_5\langle B|+_9\langle B|+\langle C|+\langle C,N|) 
e^{-lH}
 (|B\rangle_5+|B\rangle_9 +|C\rangle+|C,N \rangle),
\een
then three new amplitudes
\ben
\langle C,N| e^{-lH}|C,N\rangle,
\quad 
2 _9\langle B| 
e^{-lH}|C,N\rangle, \quad 
2 _5\langle B| e^{-lH}
|C,N\rangle .
\een
do not possess geometric interpretation.
This is a general situation in $Z_N(N=$even) model.
Thus we conclude that 
the amplitude does not factorize in 
$Z_N(N=$ even) model.
\section{Volume dependence of boundary and 
crosscap states}

We consider zero modes of boundary  and 
crosscap states.
Boundary state $|b,0,mv_i\rangle^i$ is closed string so 
that it has momentum 
and winding in compactified space.
The $m$-twisted boundary state $|b,0,mv_i
\rangle^i(mv_i\ne 0$ 
mod $N)$ sits at fixed point so that it does not have momentum 
and winding.
Hence boundary state which have momentum or winding is 
only $|b,0,0($mod$ N)\rangle^i$.
When a cylinder has Dirichlet boundary condition in $X^8 $ and $X^9$ 
directions, 
boundary closed string state can move to these directions 
and have momentum.
When a cylinder has Neumann condition in $X^8$ and 
$X^9$ directions, 
open string can move to these directions and make loop in 
compactified direction. 
Hence boundary state have winding to these directions.
Therefore $|b,0,0\rangle^i_{DD}$ has momentum
 and $|b,0,0\rangle^i_{NN}$
 has winding.
The crosscap state $|c,2kv_i,2mv_i\rangle^i$ can also have 
momentum and winding 
in compactified space.
By the same argument, the crosscap state
 $|c,2kv_i,2mv_i\rangle^i$
has momentum or winding only if
$2kv_i\equiv 0\pmod{N}$ and $2mv_i\equiv 0 \pmod{N}$.
To determine momentum and winding of $|c,0,0\rangle^i$,
we consider M\"{o}bius strip amplitude $_{NN}
{^i\langle b,0,0|}
e^{-lH}|c,0,0($mod$ 2N)\rangle^i$.
Because $_{NN}{^i\langle b,0,0|}$ has winding, 
consistency requires
 that $|c,0,0($mod$ 2N)\rangle^i$ has winding.
In the same way $|c,0,N\rangle^i$ has momentum 
because $|b,0,N\rangle^i_{DD}$ has momentum. 
Properties of the zero modes for
$|c,N($mod$ 2N),0($mod$ 2N)\rangle^i$
and $|c,N($mod$ 2N),N($mod$ 2N)\rangle^i$ 
can not be obtained
because 
$^i\langle c,N,0| e^{-lH}|c,0,0\rangle^i$
 and $^i\langle c,N,N| e^{-lH}|c,0,N\rangle^i$ vanish.

Next we consider the volume dependence produced by momentum 
and winding.
Hamiltonian which represents momentum and winding in 
compactified six dimensions is
\ben
H_2=\pi(\frac{P}{2}-L)^2,\hspace{1cm} \tilde{H}_2=\pi(\frac{P}{2}+L)^2.
\een
We denote boundary states which has momentum and winding as
\be
|b,0,0\rangle^i = |b,0,0;n,m\rangle^i=
e^{i(\frac{n\cdot x}{R_i}+m\cdot xR_i)} 
|n=0,m=0\rangle^i,\no
|c,0,0\rangle^i = |c,0,0;n,m\rangle^i=
e^{i(\frac{n\cdot x}{2R_i}+\frac{m\cdot xR_i}{2})} 
|n=0,m=0\rangle^i,
\ee
where 
$R_i$ is a common radius of the $i$ th compactified space. 
From these formula, we obtain
the volume dependence
\be
_{DD}{^i\langle b,0,0; n,0|e^{-l(H_2+\tilde{H}_2)}
|b,0,0; n,0\rangle^i}_{DD}
&=&e^{-l\pi\frac{n^2}{2R_i^2}}4\pi^2 V_i,\no
_{DD}{^i\langle b,0,0; n,0|e^{-l(H_2+\tilde{H}_2)}
|c,0,N; n,0 \rangle^i}
&=& e^{-l\pi\frac{n^2}{4R_i^2}}8\pi ^2V_i,\no
^i\langle c,0,N; n,0|e^{-l(H_2+\tilde{H}_2)}
|c,0,N; n,0 \rangle^i
&=& e^{-l\pi\frac{n^2}{8R_i^2}}16\pi ^2V_i,\no
_{NN}{^i\langle b,0,0;0,m|e^{-l(H_2+\tilde{H_2})}
|b,0,0;0, m \rangle^i}_{NN}
&=&e^{-l\pi 2m^2R_i^2}\frac{4\pi^2}{V_i},\no
_{NN}{^i\langle b,0,0;0,m|e^{-l(H_2+\tilde{H_2})}
|c,0,0;0, m \rangle^i}
&=&e^{-l\pi m^2R_i^2}\frac{8\pi^2}{V_i},\no
^i\langle c,0,0;0,m|e^{-l(H_2+\tilde{H_2})}
|c,0,0;0, m \rangle^i
&=& e^{-l\pi \frac{m^2}{2} R_i^2}\frac{16\pi^2}{V_i},
\ee
where $V_i=(R_i)^2$.
We define boundary and crosscap states $||B\rangle \rangle $ 
and $||C\rangle \rangle$ such that they include the
volume dependence of amplitudes by zero modes:
\be
||B\rangle \rangle_p
&=&\sum_{m=0}^{N-1}
\Big[N^c_{b,0,m,p}|b,0,m\rangle_p(Tr\gamma_{m,p})
\prod_{i}\frac{2\pi}{\sqrt{V_i}}\prod_{j}2\pi\sqrt{V_j}\,\Big],
\no
||C\rangle \rangle
&=&\sum_{m=0}^{N-1}\Big[N_{c,0,2m}|c,0,2m\rangle
\prod_{i}\frac{4\pi}{\sqrt{V_i}}
\prod_{j}4\pi\sqrt{V_j}\,\Big],\no
||C,N\rangle \rangle
&=&\sum_{m=0}^{N-1}
\Big[N_{c,N,2m}|c,N,2m\rangle
\prod_{i}\frac{4\pi}{\sqrt{V_i}}
\prod_{j}4\pi\sqrt{V_j}\,\Big].\label{vol}
\ee
Here $i$ and $j$ denote compactified complex planes where the states 
have momentum and winding, respectively.

By using volume dependence 
we discuss why tadpole divergence does not 
cancell in $Z_4, Z_8, Z'_8$ and $ Z'_{12}$ model.
In the case of $Z_4=\frac{1}{4}(1,1,-2)$, boundary 
and crosscap states with volume dependence are 
\be
||B\rangle \rangle_9 
&=& N_{b,0,0,9}\frac{(2\pi)^3}{\sqrt{V_{1}
 V_{2} V_{3}}}|b,0,0\rangle_9
(Tr \gamma_{0,9})
+N_{b,0,1,9}|b,0,1\rangle_9
(Tr \gamma_{1,9})\no
&&+ N_{b,0,2,9} (\frac{2\pi}{\sqrt{V_3}})|b,0,2\rangle_9
(Tr \gamma_{2,9})
+N_{b,0,3,9}|b,0,3\rangle_9
(Tr \gamma_{3,9}),
\no
||B\rangle \rangle_5
&=& N_{b,0,0,5}(\frac{2\pi}{\sqrt{V_3}})(2\pi)^2
\sqrt{V_{1} V_{2}}|b,0,0\rangle_5(Tr \gamma_{0,5})
+ N_{b,0,1,5}|b,0,1\rangle_5(Tr \gamma_{1,5})\no
&&
+ N_{b,0,2,5} (\frac{2\pi}{\sqrt{V_3}})|b,0,2\rangle_5
(Tr \gamma_{2,5})
+ N_{b,0,3,5}|b,0,3\rangle_5(Tr \gamma_{3,5}),\no
||C\rangle \rangle
&=& N_{c,0,0}\frac{(4\pi)^3}{\sqrt{V_{1} V_{2} V_{3}}}
|c,0,0\rangle
+ N_{c,0,2}(4\pi\sqrt{V_3})|c,0,2\rangle\no
& &+ N_{c,0,4}(4\pi)^2\sqrt
{V_{1} V_{2}}(\frac{4\pi}{\sqrt{V_3}})|c,0,4\rangle
+ N_{c,0,6}(4\pi\sqrt{V_3})|c,0,6\rangle,\no
||C,N\rangle \rangle
&=& N_{c,4,2}(4\pi\sqrt{V_3})|c,4,2\rangle
+ N_{c,4,6}(4\pi\sqrt{V_3})|c,4,6\rangle.
\ee
Here we omit $|c,4,0\rangle$ and $|c,4,4\rangle$ terms 
in $||C,N\rangle \rangle$ 
because these states do not contribute to the amplitude 
$\langle C,N|e^{-lH}|C\rangle$
by  $\langle c,4,0|| c,0,0\rangle 
=\langle c,4,4|| c,0,4\rangle= 0$.
$||C\rangle \rangle $ and $||C,N\rangle \rangle $
 have terms with volume factor $
\sqrt{V_3}$ but $||B^{c}\rangle \rangle_9$ and 
$ ||B^{c}\rangle \rangle_5$ do not.
It means that tadpole from $||C\rangle \rangle$ and 
$||C,N\rangle \rangle $can not be cancelled
 by $||B^{c}\rangle \rangle_9$ and 
$ ||B^{c}\rangle \rangle_5$ .
So that $Z_4$ model is inconsistent.
The same conclusion holds also for the models with
 orbifold group 
$Z_8=\frac{1}{8}(1,3,-4),
Z'_8=\frac{1}{8}(1,-3,2)$ and $Z'_{12}=\frac{1}{12}(1,5,-6)$.

In the case of $Z_3$ and $Z_7$ models, 
their amplitudes factorize 
so that 
tadpole cancellation condition of 
$Z_3$ and $Z_7$ models 
is reduced to 
massless tadpole cancellation condition in ref.\cite{ish}.
We consider the case with maximal gauge symmetry 
in which all 
p-branes sit at one fixed point, for example, at the origin.
From equations (\ref{Mc4}),(\ref{Mc3}) and (\ref{vol})
 we determine for $Z_3$ model 
\be
\lefteqn{||B\rangle\rangle _9+|C\rangle \rangle
=\frac{1}{4\sqrt{2}\pi^2}\frac{(2\pi)^3}{\sqrt{ V_1V_2V_3}}
\Big[|b,0,0\rangle_9(Tr\gamma_{0,9})
-32|c,0,0\rangle\Big]}\no
&&+\frac{1}{{4\sqrt{2}}\pi^2}
\Big[|b,0,1\rangle_9(Tr\gamma_{1,9})
-4|c,0,2\rangle\Big]
+\frac{1}{4\sqrt{2}\pi^2}
\Big[|b,0,2\rangle_9(Tr\gamma_{2,9})
+4|c,0,4\rangle\Big]
\ee
Using the massless tadpole 
cancellation  condition,
we get 
$Tr\gamma_{0,9}=32,
Tr\gamma_{1,9}=4$
 and 
$Tr\gamma_{2,9}=-4.$
It gives the gauge 
group $U(12)\times SO(8)$.
In the same way we obtain for the $Z_7$ model, 
$Tr\gamma_{0,9}=32$
 and 
$Tr\gamma_{m,9}=4
 (m=1\sim 6)$
which leads to the gauge group $U(4)^3\times SO(8)$.
They are the same conclusion which is obtained 
by one-loop diagrams \cite{ald}.

\section*{Acknowledgements}

We are grateful to S.Ishihara for useful comments and 
discussions. 

%
\section*{Figures}
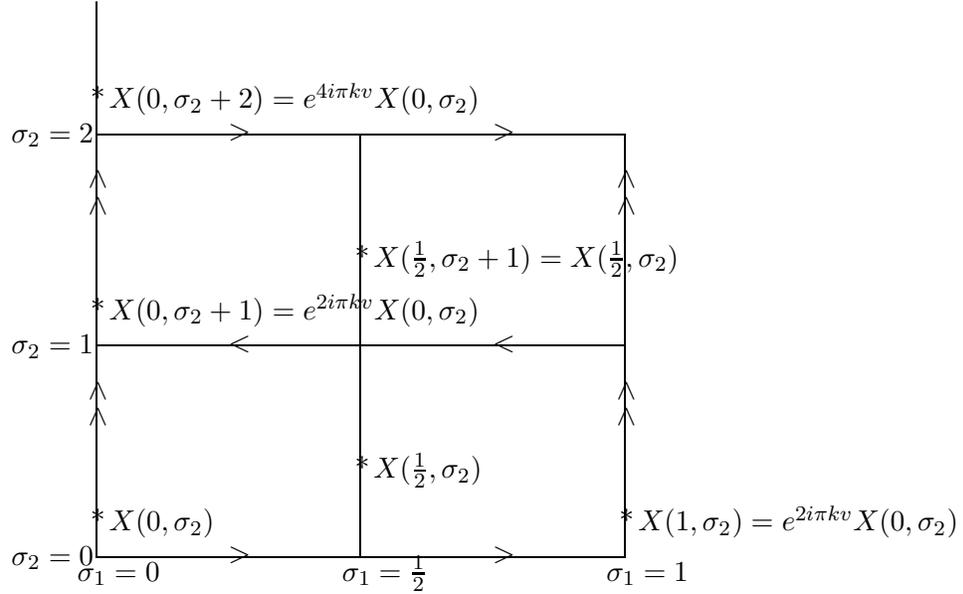
\begin{figure}[h]            
\begin{picture}(200,200)
\put(100,0){\line(2,0){200}}
\put(100,0){\line(0,2){210}}
\put(100,80){\line(2,0){200}}
\put(300,0){\line(0,2){160}}
\put(200,0){\line(0,2){160}}
\put(100,160){\line(2,0){200}}

\put(297,50){$\land$}
\put(297,60){$\land$}
\put(297,130){$\land$}
\put(297,140){$\land$}
\put(97,130){$\land$}
\put(97,140){$\land$}
\put(97,50){$\land$}
\put(97,60){$\land$}
\put(150,-2){$>$}
\put(250,-2){$>$}
\put(150,78){$<$}
\put(250,78){$<$}
\put(150,158){$>$}
\put(250,158){$>$}

\put(93,-9){$\sigma_1=0$}
\put(193,-9){$\sigma_1=\frac{1}{2}$}
\put(293,-9){$\sigma_1=1$}
\put(68,-3){$\sigma_2=0$}
\put(68,77){$\sigma_2=1$}
\put(68,157){$\sigma_2=2$}

\put(105,10){$X(0, \sigma_2)$}
\put(305,10){$X(1, \sigma_2)=e^{2i\pi kv} X(0, \sigma_2)$}
\put(105,170){$X(0, \sigma_2+2)=e^{4i\pi kv} X(0, \sigma_2)$}
\put(105,90){$X(0, \sigma_2+1)= e^{2i\pi kv} X(0, \sigma_2)$}
\put(205,30){$X(\frac{1}{2}, \sigma_2)$}
\put(205,110){$X(\frac{1}{2}, \sigma_2+1)= X(\frac{1}{2}, \sigma_2)$}

\put(98,10){*}
\put(298,10){*}
\put(98,90){*}
\put(98,170){*}
\put(198,30){*}
\put(198,110){*}

\end{picture}
\caption{Klein bottle}
\end{figure}

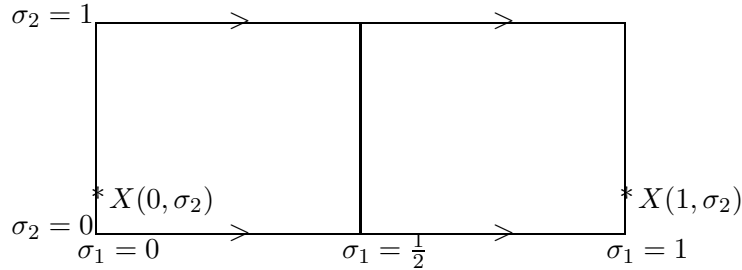
\begin{figure}[h]
\begin{picture}(200,80)
\put(100,0){\line(2,0){200}}
\put(100,0){\line(0,2){80}}
\put(100,80){\line(2,0){200}}
\put(300,0){\line(0,2){80}}
\put(200,0){\line(0,2){80}}

\put(150,-2){$>$}
\put(250,-2){$>$}
\put(150,78){$>$}
\put(250,78){$>$}

\put(93,-9){$\sigma_1=0$}
\put(193,-9){$\sigma_1=\frac{1}{2}$}
\put(293,-9){$\sigma_1=1$}
\put(68,0){$\sigma_2=0$}
\put(68,80){$\sigma_2=1$}

\put(105,10){$X(0, \sigma_2)$}
\put(305,10){$X(1, \sigma_2)$}
\put(98,10){*}
\put(298,10){*}
\end{picture}
\caption{Cylinder}
\end{figure}
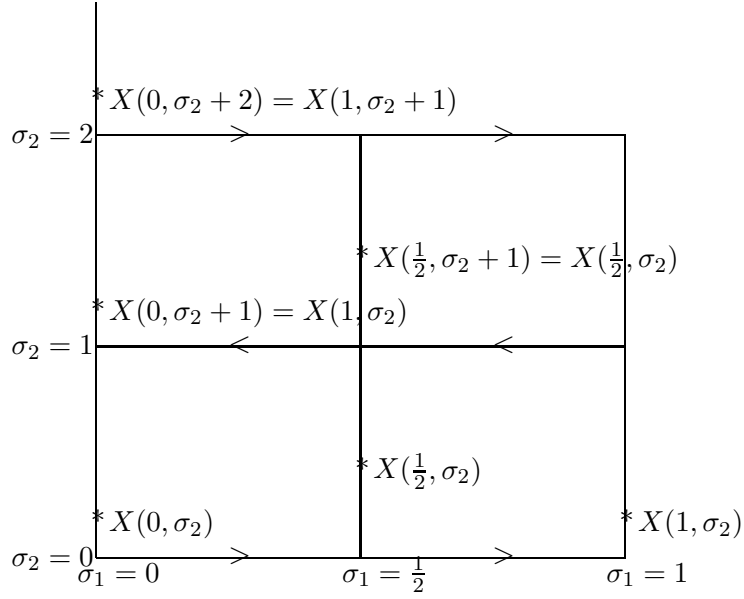
\begin{figure}[h]                      
\begin{picture}(200,180)
\put(100,0){\line(2,0){200}}
\put(100,0){\line(0,2){210}}
\put(100,80){\line(2,0){200}}
\put(300,0){\line(0,2){160}}
\put(200,0){\line(0,2){160}}
\put(100,160){\line(2,0){200}}

\put(150,-2){$>$}
\put(250,-2){$>$}
\put(150,78){$<$}
\put(250,78){$<$}
\put(150,158){$>$}
\put(250,158){$>$}

\put(93,-9){$\sigma_1=0$}
\put(193,-9){$\sigma_1=\frac{1}{2}$}
\put(293,-9){$\sigma_1=1$}
\put(68,-3){$\sigma_2=0$}
\put(68,77){$\sigma_2=1$}
\put(68,157){$\sigma_2=2$}

\put(105,10){$X(0, \sigma_2)$}
\put(305,10){$X(1, \sigma_2)$}
\put(105,170){$X(0, \sigma_2+2)=X(1, \sigma_2+1)$}
\put(105,90){$X(0, \sigma_2+1)=X(1, \sigma_2)$}
\put(205,30){$X(\frac{1}{2}, \sigma_2)$}
\put(205,110){$X(\frac{1}{2}, \sigma_2+1)= X(\frac{1}{2}, 
\sigma_2)$}

\put(98,10){*}
\put(298,10){*}
\put(98,90){*}
\put(98,170){*}
\put(198,30){*}
\put(198,110){*}

\end{picture}
\caption{M\"{o}bius strip}
\end{figure}

\begin{table}[h]
\section*{Tables}
\begin{center}
\begin{tabular}{|c|c||c|c||c|c|}  \hline
\it $Z_3$&$\frac{1}{3}(1,1,-2)$&
$Z'_6$&$\frac{1}{6}(1,-3,2)$&
$Z'_8$&$\frac{1}{8}(1,-3,2)$   \\ \hline
\it $Z_4$&$\frac{1}{4}(1,1,-2)$&
$Z'_7$&$\frac{1}{7}(1,2,-3)$&
$Z'_{12}$&$\frac{1}{12}(1,-5,4)$   \\ \hline
\it $Z_6$&$\frac{1}{6}(1,1,-2)$&
$Z_8$&$\frac{1}{8}(1,3,-4)$&
$Z'_{12}$&$\frac{1}{12}(1,5,-6)$   \\ \hline
\end{tabular}
\end{center}
\caption{$Z_N$ actions in $D=4$. Each vector stands for 
$(v_1, v_2, v_3)$. 
}
\end{table}

\end{document}